\documentclass[aps,prb,preprint,showpacs,groupedaddress]{revtex4}
\usepackage{graphicx}

\begin{document}

\title[]{Pressure induced enhancement of ferroelectricity in multiferroic  $R$Mn$_2$O$_5$($R$=Tb,Dy, and Ho)}

\author{C. R. dela Cruz$^1$, B. Lorenz$^1$, Y.Y. Sun$^1$, Y. Wang$^1$, S.Park$^2$, S-W. Cheong$^2$, M. M. Gospodinov$^3$, and C. W. Chu$^{1,4,5}$}

\affiliation{ $^1$ TCSUH and Department of Physics, University of
Houston, Houston, Texas 77204-5002, USA}

\affiliation{$^2$ Rutgers Center for Emergent Materials and Department of Physics and Astronomy,
Rutgers University, Piscataway, New Jersey 08854, USA}

\affiliation{$^3$ Institute of Solid State Physics, Bulgarian
Academy of Sciences, 1784 Sofia, Bulgaria}

\affiliation{$^4$ Lawrence Berkeley National Laboratory, 1 Cyclotron
Road, Berkeley, California 94720, USA}

\affiliation{$^5$ Hong Kong University of Science and Technology,
Hong Kong, China } %\new page

\begin{abstract}
Measurements of ferroelectric polarization and dielectric constant were done on $R$Mn$_2$O$_5$ ($R$=Tb, Dy, and Ho) with applied hydrostatic pressures of up to 18 kbar. At ambient pressure, distinctive anomalies were observed in the temperature profile of both physical properties at critical temperatures marking the onset of long range AFM order (T$_{N1}$), ferroelectricity (T$_{C1}$) as well as at temperatures when anomalous changes in the polarization, dielectric constant and spin wave commensurability have been previously reported. In particular, the step in the dielectric constant at low temperatures (T$_{C2}$), associated with both a drop in the ferroelectric polarization and an incommensurate magnetic structure, was shown to be suddenly quenched upon passing an $R$-dependent critical pressure. This was shown to correlate with the stabilization of the high ferroelectric polarization state which is coincident with the commensurate magnetic structure. The observation is suggested to be due to a pressure induced phase transition into a commensurate magnetic structure as exemplified by the pressure-temperature ($p$-$T$) phase diagrams constructed in this work. The $p$-$T$ phase diagrams are determined for all three compounds.
\end{abstract}



\maketitle

\section{Introduction}
Since the first attempts by Smolenskii et al.\cite{smolenski:59} and Venevtsev et al.\cite{venev:60} to make magnetodielectric systems in the 1950's,  many more have been discovered since the recent renaissance of the field in 2000. The revival of interest in these materials is attributed to the discovery of the coexistence and mutual interference of long range magnetic and ferroelectric orders primarily in rare earth manganites \cite{inomata:96, huang:97, lonkai:02, kagomiya:02, matsumoto:03, kimura:05}. In the case of the $R$Mn$_2$O$_5$'s ($R$=rare earth or Y) in particular, the sizable coupling between the magnetic and ferroelectric order parameters revealed by early magnetocapacitance measurements has ignited the awareness on this family of multiferroic materials\cite{chapon:04,hur:04,hur:04b,higashiyama:04}. The magnetodielectric effect can be gigantic in some and it is strong enough to trigger magnetic or dielectric phase transitions. Recent thermal expansivity measurements have shown evidence for the important role of spin-lattice interaction in the magnetodielectric coupling in these systems\cite{delacruz:06b}. The coupling between these different degrees of freedom in the material will enable one to control spontaneous magnetization (polarization) by an applied electric field (magnetic field) or stress. $Dy$Mn$_2$O$_5$, for example, has been referred to as a colossal magnetodielectric material for exhibiting more than 100 percent enhancement in its dielectric constant and allowing the control of the ferroelectric polarization upon application of a magnetic field\cite{hur:04}. Also, in $Tb$Mn$_2$O$_5$, the spontaneous polarization has been shown to be completely reversible at about 3K upon application of a magnetic field of up to 2T. With such magnetodielectric responses measured, specific device applications have been suggested for these multiferroic materials such as the multiple state memory media\cite{hur:04b}.

The $R$Mn$_2$O$_5$ compounds crystallize in the orthorhombic phase with space group Pbam at room temperature. They are composed of Mn$^{4+}$O$_6$ octahedra and Mn$^{3+}$O$_5$ bipyramids. The edge sharing linked octahedra form ribbons along the $c$-axis which are connected on the $a$-$b$ plane by the edge sharing bipyramids that are corner connected to the octahedra. The complex interplay of competing magnetic interactions between the Mn spins as well as with the rare earth moment results in complex magnetic structures characterized by a highly frustrated spin system\cite{kobayashi:04b,blake:05,ratcliff:05}. Upon cooling, the compounds undergo a cascade of magnetic and ferroelectric phase transitions suggested to be due to the inherent magnetic frustration in these systems and a competition of different states with comparable energies. From the high temperature paraelectric and paramagnetic (PM/PE) state, long range incommensurate magnetic order (HT-ICM/PE) ensues at T$_{N1}$$\approx$40-43K, described by the propagation vector \textbf{q}=(1/2+$\delta$,0,1/4+$\beta$) where $\delta$ and $\beta$ are incommensurability parameters that depend on $R$. Further cooling to T$_{C1}$$\approx$38-40K results in a lock-in transition of the magnetic ordering wavevector into a commensurate type given by \textbf{q}=(1/2,0,1/4). Moreover, the transition into a commensurate magnetic structure coincides with the onset of ferroelectric polarization along the crystal's b-axis (CM/FE1)\cite{hur:04b,higashiyama:04}. At a much lower temperature T$_{C2}$$\approx$13-18K, another sharp change in the polarization is observed which coincides with the unlocking of the commensurate order into another incommensurate magnetic order (LT-ICM/FE2). This low temperature phase has also been associated with the existence of a new excitation referred to as electromagnons \cite{sushkov:07}. Finally the rare earth moments order at T$_{C(R)}$$<$10K. The magnetic orders for all three compounds have been partially determined by neutron diffraction measurements\cite{blake:05}. These revealed that the magnetic structures of the three materials are similar in the $a$-$b$ plane regardless of the rare earth ion present. Pronounced anomalies in the dielectric constant, polarization and specific heat have been observed at each phase transition point\cite{hur:04,higashiyama:04,delacruz:06d}. Additional phase transitions have been observed for for $Ho$Mn$_2$O$_5$ and $Dy$Mn$_2$O$_5$ associated with subtle changes in the spin alignment of the magnetic structure, presumably due to the interaction with the particular rare earth moment \cite{ratcliff:05,delacruz:06d}. More importantly, however subtle these magnetic structural changes are, they are consistently reflected in distinctive anomalies in dielectric measurements exemplifying the strong magnetodielectric coupling in these multiferroic compounds.

Previous reports have highlighted the occurrence of multiple nearly degenerate magnetic ground states in these complex compounds making them highly susceptible to perturbations such as applied magnetic fields. Strong magnetic field effects observed on the magnetic structures as well as the lattice are also detected in the associated ferroelectric sub-systems \cite{higashiyama:05,ratcliff:05,blake:05}.

In this work, as opposed to coupling directly to the spin structure via application of a magnetic field, isotropic pressure was applied to the system to tune the inter-atomic distances and consequently the magnetic exchange interactions between the Mn spins. Dramatic effects of pressure have since been observed on the magnetic and ferroelectric phases of Ni$_3$V$_2$O$_8$\cite{chaudhury:07}. In this report we show the pressure induced stabilization of the ferroelectric state in $R$Mn$_2$O$_5$'s ($R$=Tb, Dy and Ho) at low temperatures. In addition, the pressure-temperature ($p$-$T$) phase diagrams were constructed for the three compounds.

\section{Methodology}
    Single crystals of $R$Mn$_2$O$_5$ ($R$=Ho,Dy,Tb) were grown via the high temperature flux growth method in solution as discussed
elsewhere \cite{hur:04,mihailova:05}. The crystals were then thinned down to $\approx$0.2-0.8 mm along the $b$-axis with gold evaporated at the surfaces to form a capacitor for the dielectric constant ($\varepsilon$$_{b}$) and ferroelectric polarization (\textbf{P$_b$}) measurements. The initial characterization of the multiferroic properties of the samples was done by determining the temperature dependence of $\varepsilon$$_{b}$ and \textbf{P$_b$} with and without an applied magnetic field and was then compared with known and published results. The temperature profile of the dielectric constant was determined from capacitance measurements using the AH2500A ultra-precision capacitance bridge (Andeen-Hagerling) at 1kHz while \textbf{P$_b$} was calculated from the measured pyroelectric current through the ammeter function of a Keithley 6517A Electrometer. In this study, the pyroelectric current was measured such that a poling field was applied throughout the measurement temperature range to accommodate for the possibility of re-entrant paraelectricity at low temperatures. The poling voltage was typically 100V corresponding to an electric field of 130-500 kV/m across the electrodes. Measurements were taken while the temperature was varied at the rate of $~$1 K/min. The measurements were also done with both forward and reverse bias voltages to check for the reversibility of the ferroelectric polarization. Subsequent measurements were done under isotropic pressures up to 18 kbar using the clamp cell method\cite{chu:74}. The pressure was measured in-situ by evaluating the change in the superconducting transition of high purity lead as affected by the applied pressure\cite{chu:74,chu:67}.

\section{Results and Discussion}

\subsection{Dielectric constant measurements under isotropic pressure}

	The existence of strong spin-lattice coupling in multiferroic $R$Mn$_2$O$_5$'s exemplified by the anomalies in the thermal expansivities and magnetostrictive coefficients reported earlier, implies a possibility of a high sensitivity of the phase transitions and physical properties to lattice strain\cite{delacruz:06d,delacruz:06b}. The strong effects of hydrostatic pressure is displayed clearly in Figures 1A, 2A and B and 3A and B showing the dielectric constant measurements done under isotropic pressure for the three rare earth manganites $R$Mn$_2$O$_5$ (R=Ho, Tb and Dy). From these dielectric measurements, pressure-temperature ($p$-$T$) phase diagrams were then constructed and will be discussed later.

There are three phase transitions that are common to all multiferroic compounds in this study. The onset of long range incommensurate AFM magnetic order at T$_{N1}$ marks a subtle break in slope at $~$43K which is seen clearly if the temperature derivative of $\varepsilon$$_{b}$(T) was taken as shown for example in the inset to Figure 1. The lock-in transition into CM/FE1 appears as a sharp peak in $\varepsilon$$_{b}$(T) and the loss factor at T$_{C1}$ (Figure 1A,2B,3B). The third common phase transition to the three compounds is at T$_{C2}$, corresponding to the unlocking of the commensurate magnetic order and is identified in $\varepsilon$$_{b}$(T) as a step-increase upon cooling. The hysteretic behavior of the ferroelectric phase transitions were also observed in accordance with previous reports.

\begin{figure}
\begin{center}
\includegraphics[width=3.3in]{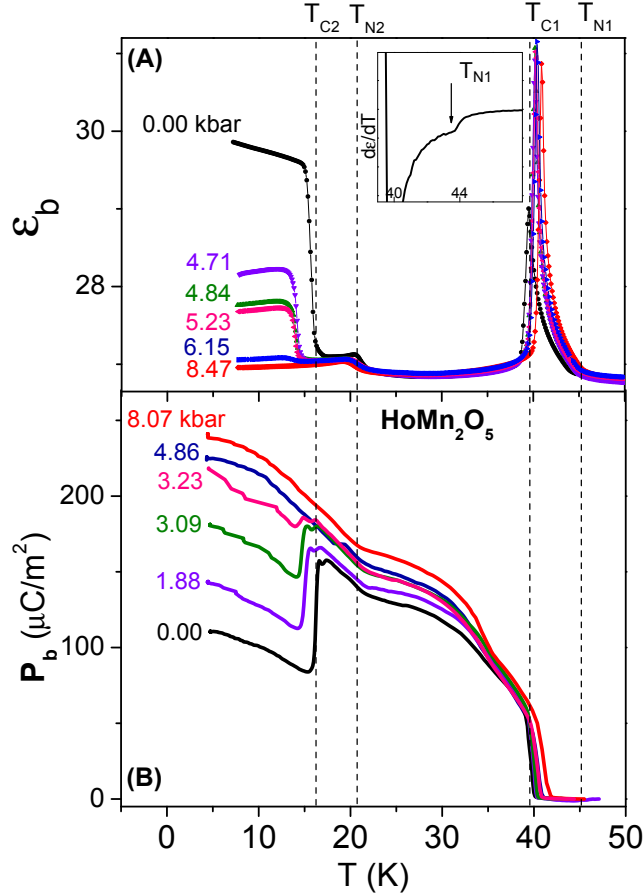}
\end{center}
\caption{Temperature profile of the (A) dielectric constant and (B)ferroelectric polarization for $Ho$Mn$_2$O$_5$ under isotropic pressures (warming only). The stabilization of the ferroelectric state is shown to occur above $p$$_C$$\approx$5kbar }
\end{figure}

Additional anomalies in the form of sharp changes in the dielectric constant are seen for both $Dy$Mn$_2$O$_5$ and $Ho$Mn$_2$O$_5$. The Mn spin system in $Ho$Mn$_2$O$_5$ has been shown to undergo a spin-reorientation at T$_{N2}$\cite{delacruz:06d}. Upon monitoring this phase transition with increasing pressure, a distinct decrease in T$_{N2}$ was observed without changing the general feature of the small step anomaly associated with it in $\varepsilon$$_{b}$(T). On the other hand, a remarkable pressure effect was seen on the sharp step anomaly of $\varepsilon$$_{b}$(T) at T$_{C2}$. In addition to the decrease of the critical temperature of the phase transition as the pressure was increased, the height of the step decreases rapidly within a limited pressure range. This behavior progresses until full quenching of the step at a critical pressure $p$$_C$$\approx$5kbar (Figure 1A). The quenching of this step, which indicates the transition into the LT-ICM phase, implies that the said phase becomes unstable above the critical pressure. We therefore propose that the commensurate magnetic order corresponding to the ferroelectric state (CM/FE1) is extended all the way to low temperatures above $p$$_C$.

\begin{figure}
\begin{center}
\includegraphics[width=3.3in]{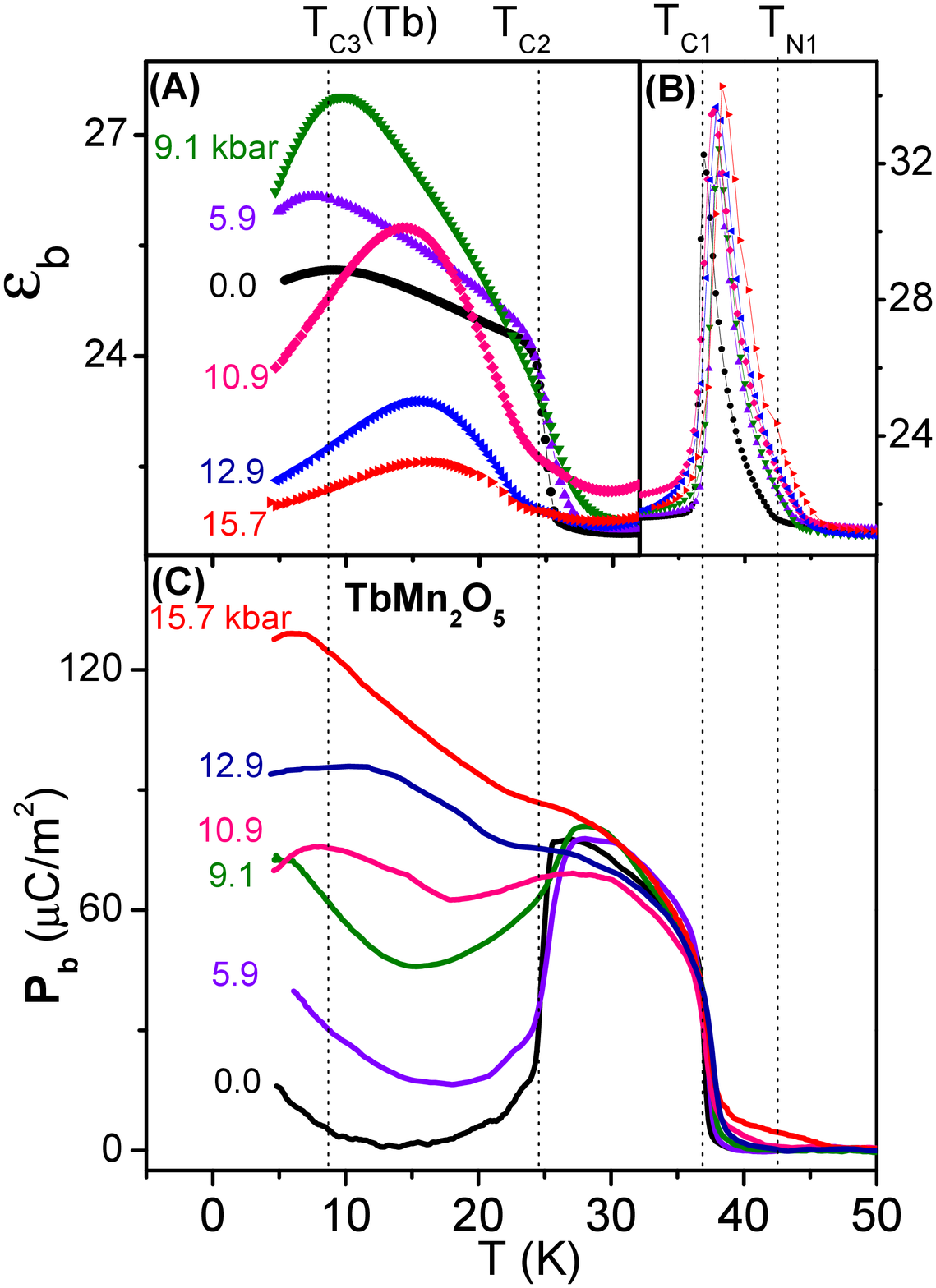}
\end{center}
\caption{Temperature profile of the (A)low temperature and (B)high temperature dielectric constant and (C)ferroelectric polarization for $Tb$Mn$_2$O$_5$ under isotropic pressures (warming only). }
\end{figure}

From the dielectric constant measurement in $Tb$Mn$_2$O$_5$, the spin rotation transition at T$_{N2}$ observed in $Ho$Mn$_2$O$_5$ appears to be missing. However, from a previous discussion, it was suggested that this coincides with the CM$\rightarrow$LT-ICM phase transition at T$_{C2}$\cite{delacruz:06d}. The phase transition at T$_{C2}$ corresponds to a sharp jump in the dielectric constant (Figure 2A). With the increase of isotropic pressure, it was observed that the magnitude of this jump rapidly decreased as well, in a similar fashion as was detected in $Ho$Mn$_2$O$_5$ (Figure 1A).
In the low temperature regime (T$<$T$_{C2}$), the temperature profile of the dielectric constant shows a broad feature (hump in $\varepsilon$$_{b}$(T)) that shifts to higher temperatures as the applied pressure was increased. This feature at low temperature has been suggested to be associated with the ordering of the Tb moments\cite{hur:04b}. Therefore, particular attention was given in distinguishing the step-like anomaly associated with the unlocking of the CM magnetic structure into the LT-ICM phase from the critical temperature associated with the hump, T$_{C3}$(Tb).

\begin{figure}
\begin{center}
\includegraphics[width=3.3in]{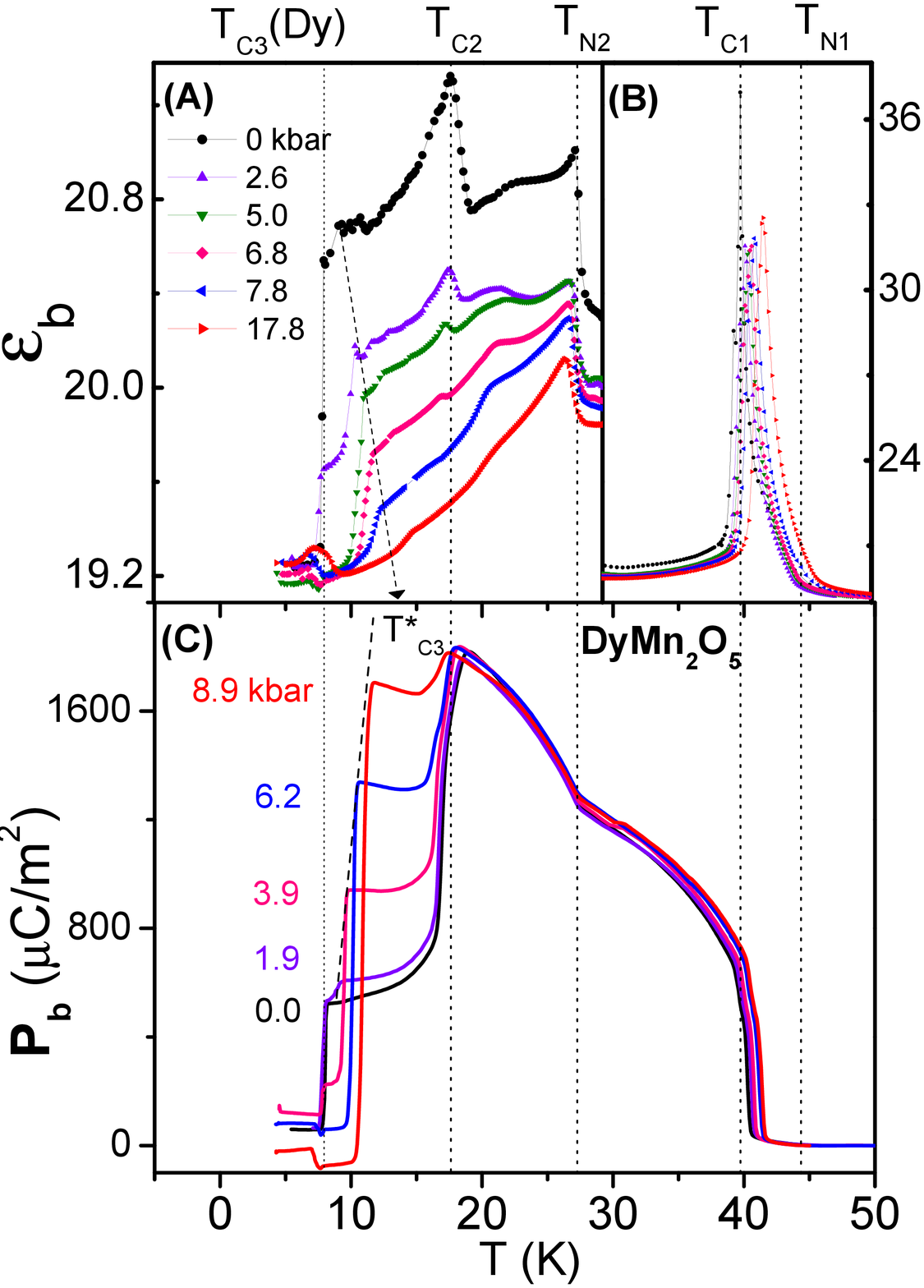}
\end{center}
\caption{Temperature profile of the (A)low temperature and (B)high temperature dielectric constant and (C)ferroelectric polarization for $Dy$Mn$_2$O$_5$ under isotropic pressures (warming only). }
\end{figure}

For the case of $Dy$Mn$_2$O$_5$, the dielectric measurement as a function of temperature taken with increasing isotropic pressure is shown in Figures 3(A and B). The high temperature phase transitions at T$_{N1}$ and T$_{C1}$ shows the same behavior with respect to pressure as that found for the other two compounds. At low temperature, the spin reorientation at T$_{N2}$ shows weak dependence on the applied isotropic pressure similar to what has been observed in $Ho$Mn$_2$O$_5$. Below the spin reorientation at T$_{N2}$,  there are two anomalies in the dielectric constant at T$_{C2}$ and T$_{C3}$(Dy). The sharp step-up at T$_{C2}$ corresponding to the CM/FE2$\rightarrow$ LT-ICM/FE3 transition exhibits a similar rapid decrease and full quenching at $p$$_C$$\approx$7kbar as was observed in the other two systems studied. Below 10K, at T$_{C3}(Dy)$ when the Dy moments order, the dielectric constant shows a sudden drop and in addition, was found to split into two anomalies upon application of pressure (Figure 3A). Also, the step at T$_{C3}$(Dy) decreases and disappears at 4kbar while the anomaly that breaks off, marked by T$^{*}$$_{C3}$  persists up to 18kbar and shifts to higher temperatures with increasing applied pressure. The splitting of the phase transition into two at T$_{C3}$ and T$^{*}$$_{C3}$ is marked by an arrow and a broken line, respectively, in Figure 3A.

\subsection{Pressure effect on the Ferroelectric Polarization of $R$Mn$_2$O$_5$ (R=Ho and Tb)}

	In the previous sections we have suggested that the LT-ICM/FE2 becomes unstable above the $R$-dependent critical pressure $p$$_C$. Since this phase is associated with a decrease in polarization (\textbf{P$_b$}), the pressure effect on its ferroelectric polarization is of primary interest. Therefore, we have developed an experimental set up to measure the pyroelectric current under hydrostatic pressure which allows for the determination of the dependence of \textbf{P$_b$} on temperature and applied pressure. Though there are unifying features on the \textbf{P$_b$} measurement results between the three compounds, we shall first discuss the results for $Ho$Mn$_2$O$_5$ and $Tb$Mn$_2$O$_5$ which are similar and leave the more complicated $Dy$Mn$_2$O$_5$ later in the next section.

Figures 1B and 2C show the results of the pioneering polarization measurements under hydrostatic pressure for $Ho$Mn$_2$O$_5$ and $Tb$Mn$_2$O$_5$. Note that the ambient pressure results are in agreement with the data presented by Higashiyama et al. ($Ho$Mn$_2$O$_5$)\cite{higashiyama:05} as well as Hur et al. and Inomata et al. ($Tb$Mn$_2$O$_5$)\cite{hur:04b,inomata:96}. In addition, the critical temperatures determined from \textbf{P$_b$}(T) and its pressure dependence agrees well with that derived from the dielectric measurement. The sharp increase of the spontaneous polarization (\textbf{P$_b$}) at T$_{C1}$ indicates the onset of ferroelectricity. Also, it is worthy to note that $Ho$Mn$_2$O$_5$ shows a subtle enhancement in \textbf{P$_b$}(T) at T$_{N2}$ signifying that the Mn spin re-orientation at that point correlates with the lattice such that the re-oriented magnetic order allows for a slight increase in the  ionic displacement resulting in an increase of \textbf{P$_b$}. The anomaly at T$_{N2}$ further proves the close correlation of the ferroelectric and magnetic orders in this system.

The sharp drop of \textbf{P$_b$} at T$_{C2}$ indicates a phase transition into the LT-ICM/FE2 phase. Since in both $Ho$Mn$_2$O$_5$ and $Tb$Mn$_2$O$_5$ the polarization remains finite, this phase is considered a weak ferroelectric phase (FE2) with a much reduced \textbf{P$_b$}. The application of pressure dramatically enhances \textbf{P$_b$}  in the FE2 phase such that the step-like drop in the polarization at T$_{C2}$ is rapidly reduced. This pressure effect is extraordinarily large for $Tb$Mn$_2$O$_5$ which exhibits a thirteenfold increase of \textbf{P$_b$} at 15 K. The anomaly at T$_{C2}$ disappears completely above the critical pressures $p$$_C$ indicating a transition into the CM/FE1 phase.

The magnitude of the change in polarization upon the transition CM/FE1$\rightarrow$LT-ICM/FE2 is different for the two compounds. $Tb$Mn$_2$O$_5$ shows a large drop of its \textbf{P$_b$} at T$_{C2}$ to a small positive value followed by a small continuous increase at lower temperatures. On the other hand, $Ho$Mn$_2$O$_5$ shows a similar sharp drop in \textbf{P$_b$}, about half its maximum value that increases to a slightly higher value at lower temperatures. For both compounds, the step-height of the sharp drop in \textbf{P$_b$}(T) at T$_{C2}$ was observed to rapidly decrease with approximately the same rate as the quenching of the step in $\varepsilon$$_{b}$(T) at the same phase transition. With the quenching of the phase transition at T$_{C2}$, the CM/FE1 state is maintained resulting in the stabilization of the high ferroelectric \textbf{P$_b$} state to low temperatures by isotropic pressure.

The stabilization of the high ferroelectric \textbf{P$_b$} state by pressure suggests a pressure induced magnetic phase transition ICM$\rightarrow$CM, stabilizing a commensurate magnetic order possibly characterized by a larger displacement of the ions which results in a higher value of the polarization \textbf{P$_b$}. In addition, this stabilization of the CM/FE1 phase by isotropic pressure is reflected by the increase of T$_{C1}$ with pressure as well as the increase of the magnitude of the ferroelectric \textbf{P$_b$} below T$_{C2}$. Notice that the applied pressure significantly increases the value of the polarization in the whole stability region of the LT-ICM/FE2 phase as well, until the step disappears at the critical pressure $p$$_C$ (Figure 1B and 2C).

A similar suppression of the step in $\varepsilon$$_{b}$(T) has been observed in $Ho$Mn$_2$O$_5$ and $Dy$Mn$_2$O$_5$ upon application of large magnetic fields along the compound's easy axis \cite{higashiyama:04,higashiyama:05,hkimura:06}. Full quenching of the step occurred at 14T for $Ho$Mn$_2$O$_5$\cite{higashiyama:05}. This observation has been associated with the magnetic field-induced transition of the magnetic order from incommensurate to commensurate, such that the ferroelectricity is stabilized. The suggestion that the magnetic field stabilized the CM/FE1 phase was confirmed by neutron scattering experiments\cite{hkimura:06}. In analogy, we propose that isotropic pressure changes the incommensurate magnetic modulation towards a commensurate one above a critical value. Neutron scattering experiments under hydrostatic pressure is thus recommended to confirm this suggestion. In addition, the reported magnetic field effect and our suggested pressure induced magnetic phase transition is in line with the recent proposal which attributes the ferroelectricity in the $R$Mn$_2$O$_5$ systems to originate from a symmetric (\textbf{S$_i$}$\bullet$\textbf{S$_j$})-type interaction referred to as "(super)exchange striction"\cite{kimura:06,cheong:07}. Another possible mechanism that has been put forth to explain the ferroelectricity in the $R$Mn$_2$O$_5$ compounds is the superposition of two ascentric spin density waves resulting to the loss of inversion symmetry\cite{chapon:06}. In both cases, the ferroelectric order parameter is sensitive to changes in the magnetic ordering wavevector, that changes with the application of magnetic fields or pressure.

\subsection{Pressure effect on the Ferroelectric Polarization of $Dy$Mn$_2$O$_5$}

The case of  $Dy$Mn$_2$O$_5$ is discussed separately because of the peculiarities of the results observed owing to its more complicated magnetic structure and its evolution with varying temperature. Temperature dependent high resolution single crystal neutron scattering experiments have revealed that $Dy$Mn$_2$O$_5$ stabilizes into multi-\textbf{k} magnetic structures where there is a coexistence of commensurate and incommensurate modulations with two distinct incommensurate orders observed\cite{ratcliff:05}. The warming data for the ferroelectric \textbf{P$_b$} taken at ambient pressure is consistent with previous reports\cite{higashiyama:04}, showing the sharp onset of ferroelectricity at T$_{C1}$ with a similar enhancement in \textbf{P$_b$} as in $Ho$Mn$_2$O$_5$ upon the spin reorientation phase transition at T$_{N2}$ (Figure 3C). Sharp drops in \textbf{P$_b$} are then seen consecutively upon crossing T$_{C2}$ and T$_{C3}$ with the major change occurring at T$_{C2}$. In correlation with the determined magnetic structures, the high \textbf{P$_b$} phase at T$_{C2}$$<$T$<$T$_{N2}$ is referred to in this work as the CM/FE2 phase, the sharp drop to the weaker ferroelectric phase at  T$_{C3}$$<$T$<$T$_{C2}$ changes the structure to a purely incommensurate one (LT-ICM/FE3) and the final drop to a re-entrant paraelectric phase at T$<$T$_{C3}$ has an associated magnetic structure that is characterized by the CM order of the Dy moments (CM(AFM-Dy)/FE4)\cite{ratcliff:05}. Upon application of pressure, the remarkable observation of a sizable increase in \textbf{P$_b$} with pressure is seen in the purely incommensurate (ambient only) phase of LT-ICM/FE3. The critical pressure is estimated to be $p$$_C$ $\approx$ 10kbar. The pressure induced stabilization of the ferroelectric high polarization state observed in both $Ho$Mn$_2$O$_5$ and $Tb$Mn$_2$O$_5$ is observed in $Dy$Mn$_2$O$_5$ strictly in the LT-ICM/FE3 phase only because the \textbf{P$_b$} is driven to a very low value at low temperatures mainly upon the onset of commensurate order of the Dy moments. In addition, the splitting of the low temperature anomaly at T$_{C3}$ seen in the dielectric measurements may also be tracked from the \textbf{P$_b$} profile. Our more detailed measurements show that the intermediate phase (X) at T$_{C3}$$<$T$<$T$^{*}$$_{C3}$ has a very small \textbf{P$_b$} at low pressures that decreases rapidly to zero as the pressure is increased. It is worth noting that the critical temperatures derived from the pressure dependent \textbf{P$_b$} measurements coincide with those determined from the dielectric measurements.

\subsection{$p$-$T$ Phase diagrams of $R$Mn$_2$O$_5$ (R=Ho, Tb and Dy)}

The observed pressure effect on the phases and phase transitions of the multiferroic systems studied may be summarized by constructing the Pressure vs. Temperature ($p$-$T$) phase diagrams for each compound. The phase diagrams were constructed based on the changes in the critical temperatures associated with the various magnetic and ferroelectric phase transitions in the compounds as the applied pressure was varied. The $p$-$T$ diagrams shown in Figures 4-6 were constructed from the warming data only for clarity.

The two phase boundaries defined by T$_{N1}$($p$) and T$_{C1}$($p$) in the high temperature part of the phase diagrams are common to the three rare earth manganites in this study. The boundaries correspond to the phase transitions PM/PE$\rightarrow$HT-ICM/PE at T$_{N1}$ and HT-ICM/PE$\rightarrow$CM/FE1 at T$_{C1}$. Both phase boundary lines show a weak positive dependence on the applied isotropic pressure of up to 18 kbar. The major differences arise at low temperatures in the $p$-$T$ phase diagrams of the three compounds, which is in part brought about by the effect of the pressure on the order of the rare earth moments.

The striking pressure effect observed from the dielectric and polarization measurements is discerned clearly for $Ho$Mn$_2$O$_5$ in Figure 4 as the suppression of the LT-ICM/FE2 phase at $p$$_C$ $\approx$ 5 kbar signifying the pressure induced transition from a low \textbf{P$_b$} incommensurate magnetic phase to a high \textbf{P$_b$} commensurate one (LT-ICM/FE2$\rightarrow$CM/FE1).

$Tb$Mn$_2$O$_5$'s $p$-$T$ phase diagram shown in Figure 5 has similarities with that derived for $Ho$Mn$_2$O$_5$ in that the LT-ICM phase becomes unstable towards the CM phase although with a higher critical pressure of $p$$_C$ $\approx$ 9 kbar.  We have decided not to include the pressure dependence of the low temperature hump in the phase diagram for it remains unclear if this represents a sharp phase transition since only a broad maximum was observed in $\varepsilon$ and no distinctive anomalies in other physical property measurements. We suggest that the Tb moments are strongly interacting with the Mn spins that results in the polarization of the Tb moments, and manifests as a broad hump-like anomaly in the dielectric measurement.

\begin{figure}
\begin{center}
\includegraphics[height=3.3in,angle=-90]{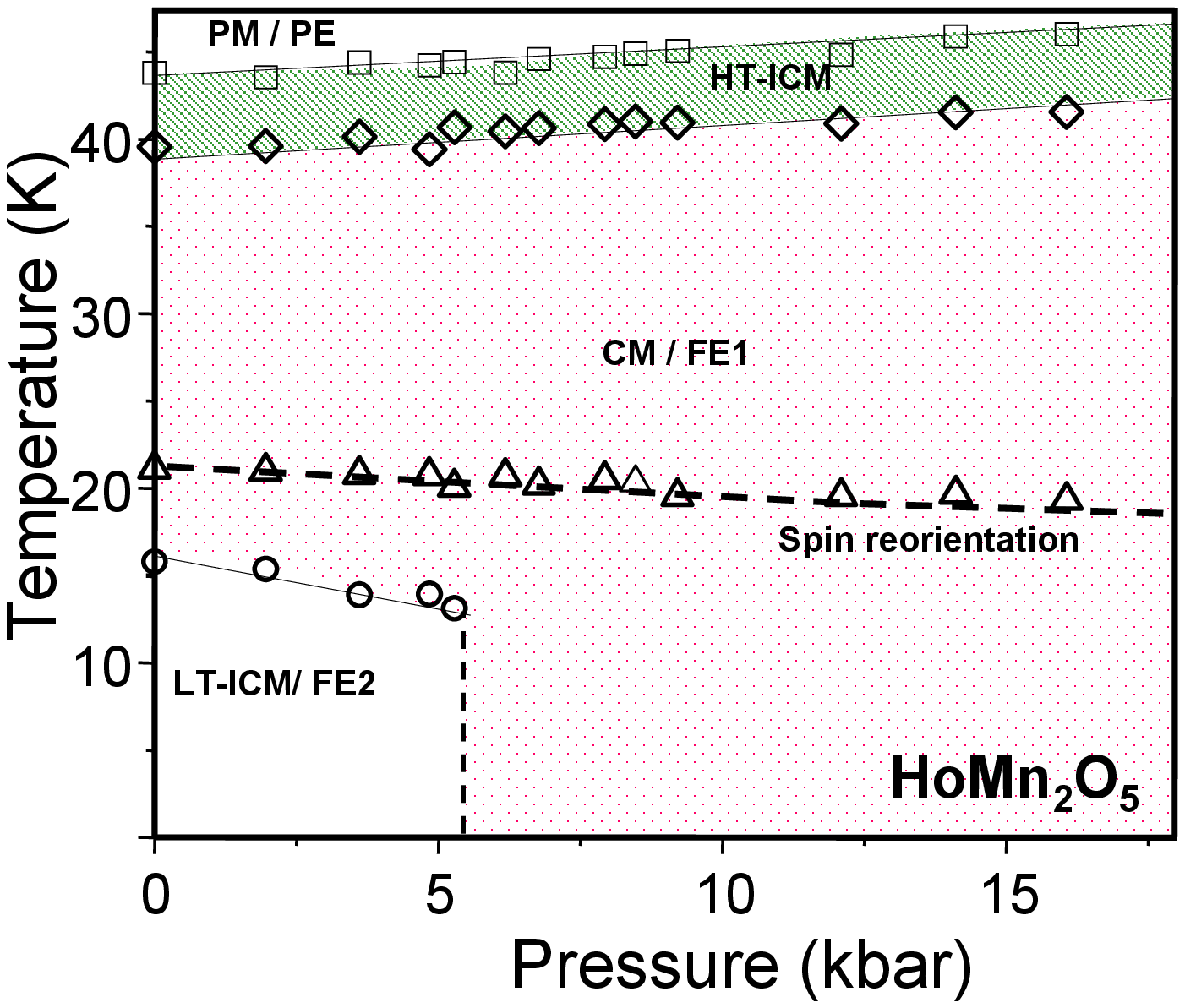}
\end{center}
\caption{Pressure-Temperature ($p$-$T$) phase diagram for $Ho$Mn$_2$O$_5$ derived from the $b$-axis dielectric measurements  with applied isotropic pressure. The shaded area represents the pressure range above which the LT-ICM/FE2 phase destabilizes into the CM/FE1 phase. }
\end{figure}

\begin{figure}
\begin{center}
\includegraphics[height=3.3in,angle=-90]{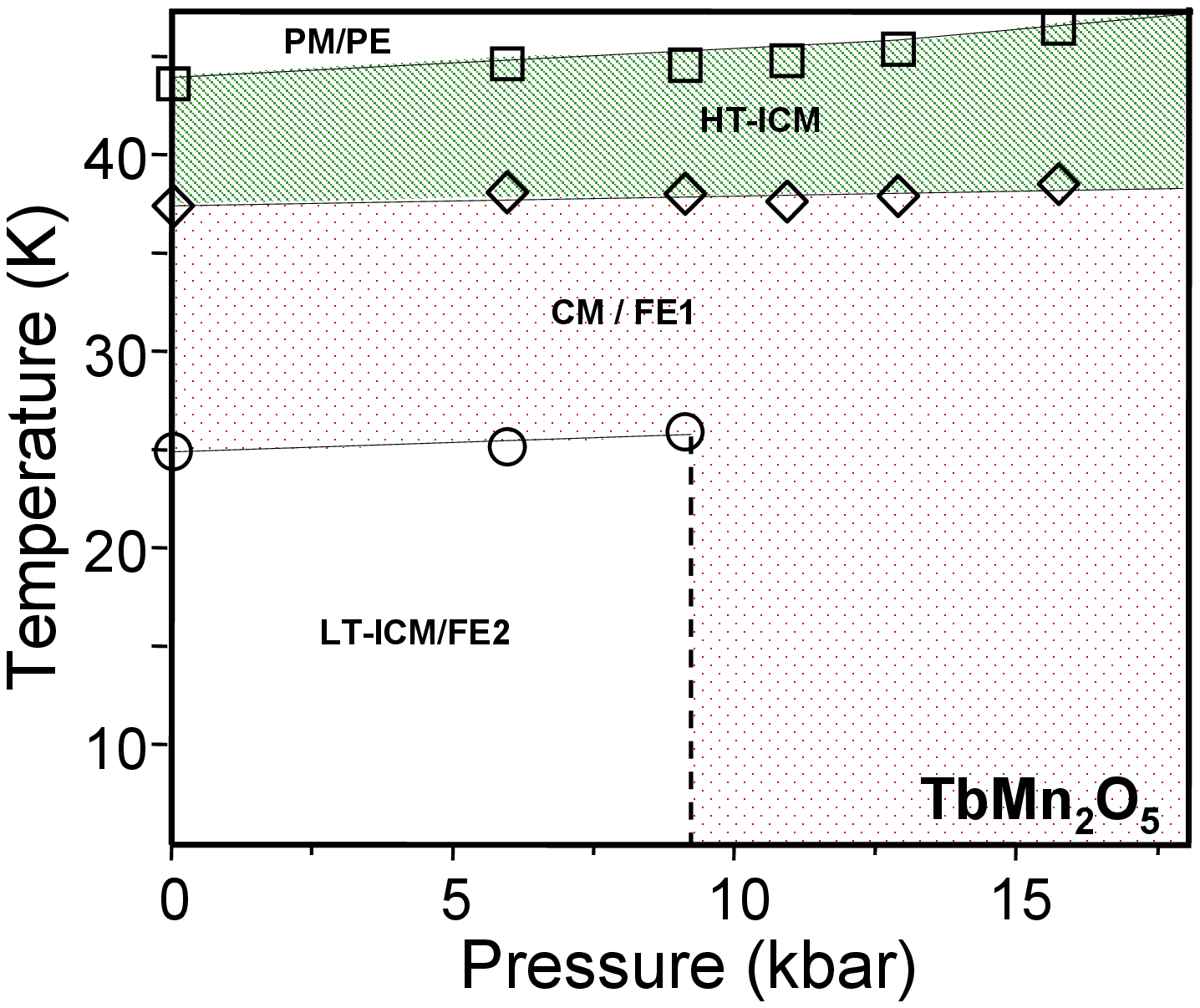}
\end{center}
\caption{Pressure-Temperature ($p$-$T$) phase diagram for $Tb$Mn$_2$O$_5$ derived from the $b$-axis dielectric measurements  with applied isotropic pressure. The shaded area represents the pressure range above which the LT-ICM/FE2 phase destabilizes into the CM/FE1 phase. }
\end{figure}

\begin{figure}
\begin{center}
\includegraphics[height=3.3in,angle=-90]{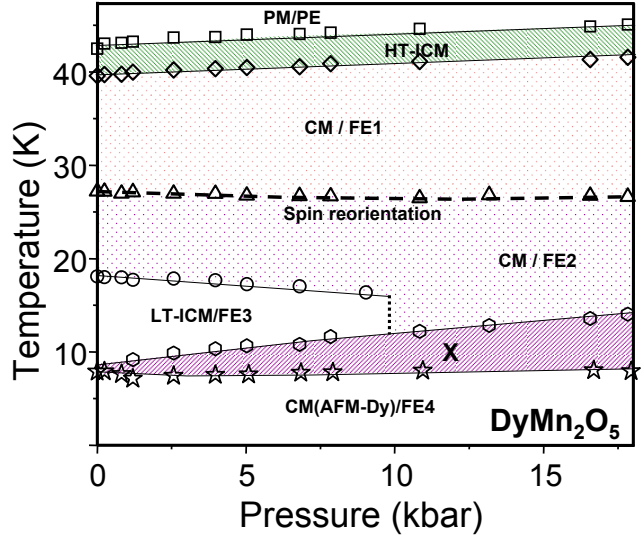}
\end{center}
\caption{Pressure-Temperature ($p$-$T$) phase diagram for $Dy$Mn$_2$O$_5$ derived from the $b$-axis dielectric measurements  with applied isotropic pressure. The pressure separated X phase was found to be paraelectric but the nature of the magnetic structure is yet to be determined.}
\end{figure}

Among the three $R$Mn$_2$O$_5$ systems studied, $Dy$Mn$_2$O$_5$ shows the most number of phase transitions at ambient pressure and the cascade of transitions evolve upon application of a magnetic field resulting to a complex magnetic phase diagram\cite{ratcliff:05}. In addition to the high temperature phase boundaries which behave similarly as in the case of the other two multiferroics, an interesting phase splitting due to pressure was observed in $Dy$Mn$_2$O$_5$. It creates an intermediate phase (X in Figure 6) at low pressures between LT-ICM/FE3 and the low temperature re-entrant PE phase (CM(AFM-Dy/FE4)). In the X phase, \textbf{P$_b$} rapidly decreases as isotropic pressure was applied. The nature of the magnetic structure of this phase is yet to be explored and is suggested to be related to the stabilization with increasing pressure of either the low temperature CM order of Dy moments, or one of the incommensurate orders. The pressure induced stabilization of ferroelectricity that has been observed in $Ho$Mn$_2$O$_5$ and $Tb$Mn$_2$O$_5$, is seen clearly in the phase diagram as a sharp phase line at $p$$_C$ $\approx$ 10 kbar delineating the purely incommensurate low \textbf{P$_b$} LT-ICM/FE3 phase from the commensurate high \textbf{P$_b$} CM/FE2 phase.

In Figures 4-6, the unshaded area at low $T$ and $p$ represents the pressure range above which the high \textbf{P$_b$} ferroelectric CM/FE1(FE2) state is stabilized or restored for $R$=Ho,Tb (Dy). In addition, the observed pressure induced enhancement of the ferroelectric \textbf{P$_b$} in the entire stability region of the LT-ICM/FE2(3) phase is such a remarkable effect and it is so far observed only in the multiferroic $R$Mn$_2$O$_5$'s.

Recent optical measurements have observed electromagnon excitations in that the LT-ICM phase. It was shown that the particular excitations were responsible for the sharp increase in the dielectric response observed at T$_{C2}$\cite{sushkov:07}. In correlation with the said results, it is suggested that an applied pressure will suppress the electromagnon excitations which may be verified by performing optical measurements under pressure. The suppression of the electromagnon signal should be in a similar manner as the observed pressure induced quenching of the low \textbf{P$_b$ }LT-ICM phase.

\section{Summary}

The dielectric constant and ferroelectric polarization were measured under hydrostatic pressure and anomalies associated with the different magnetic and ferroelectric phase transitions have been used to construct the complete $p$-$T$ phase diagram for $R$Mn$_2$O$_5$ ($R$=Ho, Tb and Dy). The resulting $p$-$T$ phase diagrams show the phase boundaries defining the different phases in all three compounds. The complex phase diagram derived for $Dy$Mn$_2$O$_5$ particularly in the low temperature region remains tentative at the moment until further characterization of the magnetic structure of the  pressure separated phases are done. More importantly, the applied pressure was found to quench the LT-ICM/FE2(FE3) phase and stabilizes the high polarization CM/FE1(FE2) phase for $R$=Ho,Tb (Dy). A dramatic enhancement of the polarization was observed under pressure in the LT-ICM/FE2(FE3) that is best seen in $Tb$Mn$_2$O$_5$ which at 12 kbar and 15 K exhibits a \textbf{P$_b$} that is about thirteen times greater than its value at ambient pressure. Neutron diffraction measurements under pressure should be done to show this pressure effect on the magnetic structure that leads to the remarkable effect on the ferroelectric order.
	It is interesting to see that the application of pressure has the same result as the application of a magnetic field along the easy axis on the ferroelectric phase stability in these compounds. While the magnetic field changes the magnetic structure by aligning the spins with the applied field, the pressure directly changes the exchange coupling constant by influencing the interatomic distances and bond angles. A deeper understanding of this field and pressure effect entails a microscopic model describing the magnetic order, the elastic forces and the coupling between them.

\begin{acknowledgments}
This work is supported in part by NSF Grant No. DMR-9804325, the
T.L.L. Temple Foundation, the J. J. and R. Moores Endowment, and the
State of Texas through the TCSUH and at LBNL by the DoE. The work of
M. M. G. is supported by the Bulgarian Science Fund, grant No F
1207. The work at Rutgers was supported by the NSF-DMR-0520471.
\end{acknowledgments}

%REFERENCES

\end{document}